%Paper: hep-ph/9211334
%From: Stamatis Vokos <vokos@galileo.phys.washington.edu>
%Date: Mon, 30 Nov 92 12:25:58 MST

\input harvmac.tex

%%%%%%%%%%%%%%%%%%%%%% temporary, until harvmac fixed:

\def\refmark#1{${}^{\refs{#1}}$}
\def\footsym{*}\def\footsymbol{}\ftno=-2
\def\foot{\ifnum\ftno<\pageno\xdef\footsymbol{}\advance\ftno by1\relax
\ifnum\ftno=\pageno\if*\footsym\def\footsym{$^\dagger$}\else\def\footsym{*}\fi
\else\def\footsym{*}\fi\global\ftno=\pageno\fi
\xdef\footsymbol{\footsym\footsymbol}\footnote{\footsymbol}}
%%%%%%%%%%%%%%%%%%%%%%%%%%%%%%%%%%%%%%%%%%%%%%%%%%%%%%%%%%%%%%%%%%%%%%%%%

\font\authorfont=cmcsc10 \ifx\answ\bigans\else scaled\magstep1\fi

\lref\weldonfp{H.\ A.\ Weldon, West Virginia University Preprint
WVU-992 (August 1992).}
\lref\weldondisc{H.\ A.\ Weldon, \sl Phys.\ Rev.\ \bf D28\rm, 2007
(1983). Our expression (4.4b) for the $T\neq 0$ cut disagrees with Weldon's
corresponding cut.}
\lref\puzzle{P.\ Arnold, E.\ Braaten, and S.\ Vokos, \sl Phys.\ Rev.\ \bf
D46\rm, 3576 (1992).}
\lref\bd{P.\ Bedaque and A.\ Das, Rochester Preprint UR 1275, ER-40685-727
(September 1992).}
\lref\jeon{
A.\ A.\ Abrikosov, L.\ P.\ Gorkov and I.\ E.\ Dzyaloshinski,
``Methods of Quantum Field Theory in Statistical Physics'', Dover, New York,
1963;\hfill\break
A.\ L.\ Fetter and J.\ D.\ Walecka, ``Quantum Theory of Many Particle
Systems'', McGraw-Hill, New York, 1971;\hfill\break
H.\ Umezawa, H.\ Matsumoto, and M.\ Tachiki, ``Thermo Field Dynamics
and Condensed States'', North-Holland, Amsterdam, 1982;\hfill\break
S.\ Jeon, Univ.\ of Washington Preprint UW/PT-92-03.}
\lref\ewpt{
M.\ Carrington, \sl Phys.\ Rev.\ \bf D45\rm, 2933 (1992);\hfill\break
M.\ Dine, G.\ Leigh, P.\ Huet, A.\ Linde, and D.\ Linde, \sl Phys.\ Rev.\
\bf D46\rm, 550 (1992), and \sl Phys.\ Lett.\ \bf B283\rm, 319 (1992);
\hfill\break
G.\ Boyd, D.\ Brahm, S.\ Hsu, Enrico Fermi Institute Preprint
EFI-92-22 (1992);\hfill\break
P.\ Arnold and O.\ Espinosa, Univ.\ of Washington Preprint UW/PT-92-18.}
\lref\analytic{
Y.\ Fujimoto and H.\ Yamada, \sl Z.\ Phys.\ \bf C37\rm, 265 (1988);\hfill\break
P.\ S.\ Gribosky and B.\ R.\ Holstein, \sl Z.\ Phys.\
 \bf C47\rm, 205 (1990);\hfill\break
P.\ Bedaque and A.\ Das, \sl Phys.\ Rev.\ \bf D45\rm,
 2906 (1992)}
\lref\therm{V.\ P.\ Silin,  \sl Sov.\ Phys.\ JETP \bf 11\rm, 1136 (1960);
\hfill\break
  D.J. Gross, R.D. Pisarski and L.G. Yaffe, \sl Rev.\ Mod.\ Phys.\ \bf 53\rm,
  43 (1981);\hfill\break
  V.V. Klimov, \sl Sov.\ Phys.\ JETP \bf 55\rm, 199 (1982);\hfill\break
  H.A. Weldon, \sl Phys.\ Rev.\ \bf D26\rm, 1394 (1982);\hfill\break
  E.\ Braaten and R.\ D.\ Pisarski, \sl Nucl.\ Phys.\ \bf B337\rm, 569 (1990)
and \bf B339\rm, 310 (1990), \sl Phys.\ Rev.\ Lett.\ \bf 64\rm, 1338 (1990),
\sl Phys.\ Rev.\ \bf D45\rm, 1827 (1992);\hfill\break
J.\ Frenkel and J.\ C.\ Taylor, \sl Nucl.\ Phys. \bf B334\rm, 199
(1990) and \bf B374\rm, 156 (1992).}
\lref\kapusta{The reader wishing an introduction to perturbation theory
  at finite temperature may try J.I. Kapusta,
  ``Finite-temperature field theory''
  Cambridge Univ.\ Press, Cambridge, 1989.}

\Title{
  \vbox{\baselineskip12pt
    \hbox{UW/PT-92-19}
    \hbox{ANL-HEP-PR-92-51}
    \hbox{UR-1287}
    \hbox{ER-40685-736}}
  }{
  \vbox{
    \centerline{On the Analytic Structure of the Self-Energy for}
    \centerline{Massive Gauge Bosons at Finite Temperature}
  }}
\centerline{\authorfont Peter Arnold}
\centerline{\authorfont Stamatis Vokos}
\centerline{\sl Department of Physics, FM-15,}
\centerline{\sl University of Washington,}
\centerline{\sl Seattle, WA 98195}
\bigskip
\centerline{\authorfont Paulo Bedaque}
\centerline{\authorfont Ashok Das}
\centerline{\sl Department of Physics and Astronomy,}
\centerline{\sl University of Rochester,}
\centerline{\sl Rochester, NY 14627}

\vskip .3in
We show that the one-loop self-energy at finite temperature has a unique
limit as the external momentum $p_\mu\rightarrow 0$ {\it if} the loop involves
propagators with distinct masses. This naturally arises in theories involving
particles with different masses as is demonstrated for a toy model of two
scalars as well as in a $U(1)$ Higgs theory. We
show that, in spontaneously broken gauge theories, this observation
nonetheless does not affect the difference between the Debye and
plasmon masses, which are often thought of as the $(p_0=0, \vec{p}\to
0)$ and $(p_0\to 0,\vec{p}=0)$ limits of the self-energy.

\Date{October 1992}

%\draftmode

\newsec{Introduction}

In finite temperature field theory, the existence of an additional four-vector,
namely the four-velocity of the plasma, allows one to construct two
independent Lorentz scalars on which the Green's functions can depend,
\eqna\introa
$$
\eqalignno{
\omega &= P\cdot u\, &\introa a\cr
k &= \sqrt{[(P\cdot u)^2-P^2]}\, &\introa b\cr}
$$
where $u^\mu$ is the four-velocity of the plasma and $P^\mu=(p^0,\vec{p})$ is
the four-momentum of any particle.
In the rest-frame of the fluid, these scalars reduce to $p^0$ and
$p=|\vec{p}|$ respectively.

In particular,
this separate dependence of polarization tensors and self-energies, allows one
to take the limits $p^0\rightarrow 0$ and $p\rightarrow 0$ in different orders
and, in general, one
expects that the limits need not commute. In fact, it has been shown that
in the case of hot QCD\refmark{\therm},
and self-interacting scalars\refmark{\analytic,\weldondisc},
the two limits do not indeed commute. In this paper, however, we will
show that there exist contributions to the one-loop self-energy of a
massive gauge boson in a
spontaneously broken gauge theory which possess a unique limit as $p$ and
$p^0$ tend to zero, as long as the particles propagating in the loop
have different masses.
As we shall show, however, for the purposes of computing {\it
physical} quantities, such as poles of particle propagators, the usual
approximation which uses the non-commuting limits is perfectly adequate.

The outline of the paper is as follows. In Section~2, we study the
contribution to the self-energy of a scalar field $\phi_1$ through its
coupling to another scalar $\phi_2$, ${\cal L}_I=-{\lambda\over 2} {\phi_1}^2
\phi_2$, where the two fields have different masses.
In Section~3, we analyze the polarization tensor of the massive photon
in a spontaneously broken $U(1)$ theory, where the Higgs
and the photon have different masses.
Finally, we conclude with
comments on the physical interpretation of our results in Section~4.
For pedagogical completeness,  we include, in the appendix,
the derivation of the vector polarization tensor. In that calculation,
we employ Feynman parametrization and $\epsilon$-regularization in the real
time formalism, paying particular attention to the subtleties of Feynman
parametrization pointed out by Weldon\refmark{\weldonfp}.

\newsec{The Scalar Case}

Consider a toy model described by the Lagrangian density,
\eqn\lags{
{\cal L}={\cal L}_0^1(\phi_1)+ {\cal L}_0^2(\phi_2)-{\lambda\over 2}
 \phi_1^2\phi_2\,}
where
\eqn\lfree{
{\cal L}_0^i={1\over2}\partial_\mu\phi_i\partial^\mu\phi_i-{m^2_i\over2}
\phi_i^2,}
with $i=1,2$ (no summation).
We ignore the fact that the potential for this theory is unbounded from
below, and compute the one-loop contribution to the
self-energy of $\phi_1$ to demonstrate its analytic structure.
The only contribution to the self-energy at one-loop
comes from the diagram in \fig\scalar{The one-loop contribution to the
$\phi_1$ self-energy.}.
One obtains
the temperature-dependent part of the self-energy in the usual fashion\refmark
{\kapusta},
\eqn\sigk{
\eqalign{
{\rm Re}\Sigma^\beta (p_0,\vec{p}) =-\lambda^2\int {d^3k\over (2\pi)^3}\big[
&{n(\omega_1)\over 2\omega_1} {1\over
(P^2-2\vec{k}\cdot\vec{p}+2\omega_1 p_0+\Delta m_{12}^2)}\cr+
&{n(\omega_2)\over 2\omega_2} {1\over
(P^2+2\vec{k}\cdot\vec{p}+2\omega_2 p_0-\Delta m_{12}^2)}\cr
&+ (p_0\leftrightarrow -p_0)\big]\cr}
}
where $\omega_i^2=\vec{k}^2+m_i^2$, for $i=1,2$,
$\Delta m_{12}^2=m_1^2-m_2^2$, and
$n(x)$ is the Bose-Einstein distribution.

The angular integration yields
\eqn\final{
{\rm Re}\Sigma^\beta(p_0,\vec{p})=-{\lambda^2\over 8\pi^2}\int_0^\infty dk\,
{k\over p}
\left[{n(\omega_1)\over 2\omega_1}\ln \left|S_1\right|+
      {n(\omega_2)\over 2\omega_2}\ln \left|S_2\right|\right]\,,}
where
\eqna\ss
$$
\eqalignno{
S_1 &= {(p_0^2-p^2+\Delta m_{12}^2+2pk)^2-4p_0^2\omega_1^2\over
        (p_0^2-p^2+\Delta m_{12}^2-2pk)^2-4p_0^2\omega_1^2}\,, &\ss a\cr
S_2 &= {(p_0^2-p^2-\Delta m_{12}^2+2pk)^2-4p_0^2\omega_2^2\over
        (p_0^2-p^2-\Delta m_{12}^2-2pk)^2-4p_0^2\omega_2^2}\,. &\ss b\cr}
$$

\noindent
In order to analyze the behavior of the self-energy close to $P^\mu=0$,
let us assume $p_0=\alpha p$ in \final\ and take the limit $p\rightarrow 0$.
If the result is $\alpha$-dependent, then the value of the double
limit will depend on the way $P^\mu$ approaches zero and, therefore,
would correspond to a non-analytic structure at $P^\mu=0$. It is easy to
establish that \final\ in this limit becomes,
\eqn\limits{
\lim_{p\rightarrow 0}{\rm Re}\Sigma^\beta(\alpha p,\vec{p})=
-{\lambda^2\over \pi^2
\Delta m_{12}^2}\int_0^\infty dk\ k^2\left[{n(\omega_1)\over 2\omega_1}-
{n(\omega_2)\over 2\omega_2}\right]\,.}
This limit clearly is independent of $\alpha$, and therefore takes on a unique
value showing that the self-energy is analytic at $P^\mu=0$.
This is also the value that one would have obtained for
the self-energy  by setting $P^\mu=0$
inside the integrand in Eq.~\sigk. In such a case,
it is clear from (2.3) that $\Delta m^2_{12}$
has to be non zero for a well behaved result. At high temperature, the leading
contribution to this double limit can be shown to be
\eqn\hitemp{
\lim_{{{p\rightarrow 0}\atop {p_{\scriptscriptstyle 0}\rightarrow 0}}}
{\rm Re}\Sigma^\beta(p_0,p)\simeq {\lambda^2\over 4\pi}
{T\over m_1+m_2}\,. }

\noindent
Next, let us turn our attention to a spontaneously broken U(1) gauge theory.

\newsec{Abelian Higgs Model}

For simplicity, we will perform the calculation of the polarization
tensor for the massive vector boson in the Abelian Higgs model in
unitary gauge. Unitary gauge is
infamous for complications in the Higgs sector at
finite temperature\refmark{\puzzle}. In the gauge sector, however, these
complications are absent and the smaller number of diagrams makes its
use preferable for our purposes.

The Lagrangian for the Abelian Higgs model in the
unitary gauge is given by
\eqn\lagu{
\eqalign{{\cal L}=& -{1\over 4}F^{\mu\nu} F_{\mu\nu} + {e^2 v^2\over 2} A^\mu
A_\mu
+ {1\over 2} \partial^\mu\eta\partial_\mu\eta - {m^2\over 2} \eta ^2\cr
& + {e^2\over 2}A^\mu A_\mu\eta^2+e^2 v A^\mu A_\mu\eta -\lambda v \eta^3
-{\lambda\over 4} \eta^4 ,\cr}}
\noindent
where $\eta$ is the Higgs field, $A_\mu$ is the U(1) gauge field and
the vacuum expectation value, $v=m/\sqrt{2\lambda}$.
 In this gauge, the only one-loop, momentum-dependent correction to the photon
propagator is
given by the diagram in \fig\vector{The only one-loop,
momentum-dependent contribution to the vector self-energy, in unitary gauge.},
which we denote by
$\tilde\Pi_{\mu\nu}$. This diagram gives,
\eqn\picomp{
\eqalign{{\rm Re}\tilde\Pi_{00}^\beta=4e^2\int {d^3k\over (2\pi)^3}
\Big[
&{n(\omega_k)\over 2\omega_k}\,{M^2-(p_0-\omega_k)^2\over
(p_0-\omega_{k})^2-\Omega_{k+p}^2}
+{n(\Omega_k)\over 2\Omega_k}\,{M^2-\Omega_k^2\over
(p_0-\Omega_k)^2-\omega_{k+p}^2}\Big]\cr
+&(p_0\to -p_0)\,.\cr}}
Here we have defined $M=ev$, $\omega_k=\sqrt{\vec k^2+m^2}$ and
$\Omega_k=\sqrt{\vec k^2+M^2}$. (The expression \picomp\ is easily obtained by
standard techniques. However, because confusion in the literature
about the $P^\mu\to 0$ limit is often tied to particular techniques for
computing finite-temperature diagrams, and since the use of Feynman
parametrization in the real-time formalism is a particularly
good example of this, we show explicitly in the appendix how to
perform such a calculation using this technique,
together with regularization of the real-time propagators. For routine
calculations, however, Feynman parametrization is impractical.)
After doing the angular integration, one obtains
\eqn\pik{
\eqalign{
{\rm Re}\tilde\Pi_{00}^\beta(p_0,p)=-{e^2\over 2\pi^2}\int_0^\infty dk\, k
\Bigg[&{(\Delta m_{12}^2+k^2+p_0^2)n(\omega_k)\over2\omega_k}{1\over p}
\ln\left|S_1\right|+{k^2n(\Omega_k)\over2\Omega_k}{1\over p}\ln\left|S_2
\right|\cr
&+n(\omega_k){p_0\over p}
\ln\left| {(p_0^2-p^2+\Delta m_{12}^2)^2-4(p_0\omega_k+pk)^2\over
(p_0^2-p^2+\Delta m_{12}^2)^2-4(p_0\omega_k-pk)^2}\right|\Bigg]\,,\cr
}}
where $S_i$ are given in \ss\, with $m_1=m$ and $m_2=M$.

Let us analyze the small-$p^0$, small-$p$ behavior of \pik. For that purpose,
let us set as before
\eqn\pzerop{
p^0=\alpha p\,.}
Then, for {\it nonzero}\ values of $\Delta m_{12}^2=m^2-M^2$, it is
clear that
\eqn\pioozero{
\lim_{p\to 0}{\rm Re}\tilde\Pi_{00}^\beta(\alpha p,p)=
-{4e^2\over \pi^2}\int_0^\infty dk\left[k^2{n(\omega_k)\over 2\omega_k}+
{k^4\over m^2-M^2}\left({n(\omega_k)\over2\omega_k}-{n(\Omega_k)\over
2\Omega_k}\right)\right]\,.}
In particular, this limit is finite, $\alpha$-independent and hence
independent of the ratio $p_0/p$ as $p_0$ and $p$ approach zero.
Alternatively, this may be obtained by simply putting $P^\mu=0$ in
\picomp. So, the double limit is unique, as promised.
Furthermore, it is easy to establish that ${\rm Re}\tilde\Pi_{ii}^\beta$
has a unique double limit as well.

The high-temperature limit of \pioozero\ can be easily obtained to be
\eqn\pioohit{
\lim_{{p\to 0}\atop {p_0\to 0}}{\rm
Re}\tilde\Pi_{00}^\beta(p_0,p)={1\over 2}e^2 T^2\,,}
which turns out to be the same as the $(p_0=0,\vec{p}\to 0)$ limit of
the equal mass case $\Delta m_{12}^2=0$. (In fact, we note here that
even though the expression \pioozero\ appears to be singular when
$m=M$, it indeed has a finite limit as the two masses become
degenerate and corresponds to the $p_0=0, \vec{p}\to 0$ limit of the
degenerate case. One can, therefore, even foresee using such a mass-splitting
regularization in such calculations for the equal mass case.)

\newsec{Summary and Physical Implications}
We have shown, both in the context of a scalar toy model as well as for a
spontaneously broken Abelian gauge theory, that the finite-temperature
one-loop self-energy/polarization tensor at finite temperature
has a unique limit as the external four-momentum goes to zero.
The absence of the usual non-commuting double limits is traced to the fact
that there is (generically) a finite mass difference among the particles
propagating in the loop. One can understand this result in the following
way. The real part of the one-loop self-energy is related to the imaginary
part through the dispersion relation\refmark{\weldonfp},
\eqn\didp{
\eqalign{
{\rm Re}\Sigma^\beta_R(p_0,p)&={1\over \pi}{\cal P}\int_{-\infty}^\infty
du {{\rm Im}\Sigma^\beta_R(u,p)\over u-p_0}\cr
&={2\over\pi}{\cal P}\int_0^\infty du\,u{{\rm Im}\Sigma^\beta_R(u,p)\over
u^2-p_0^2}\,.}}
The last equality follows from the fact that ${\rm Im}
\Sigma^\beta_R(p_0,p)$
is an odd function of $p_0$\refmark{\jeon}. Here $\Sigma^\beta_R$ is the
retarded two point function related to $\Sigma^\beta$ by
\eqn\sig{
\eqalign{
{\rm Im}\Sigma^\beta_R(u,p)&={\rm Im}\Sigma^\beta(u,p)\,{\rm tanh}{\beta
u\over 2}\cr
{\rm Re}\Sigma^\beta_R(u,p)&={\rm Re}\Sigma^\beta(u,p)\, .}}

As pointed out by
Weldon\refmark{\weldondisc}, ${\rm Im}\Sigma^\beta_R(u,p)$ is
non-zero only for some values of $u^2-p^2$. The imaginary part of the
self-energy is expressed in terms of the discontinuity of $\Sigma^\beta_R
(p_0,p)$ along these cuts on the real axis,
\eqn\disc{
\lim_{\epsilon\to 0^+}\left(\Sigma^\beta_R(p_0+i\epsilon,p)
-\Sigma^\beta_R(p_0-i
\epsilon,p)\right)=-2i{\rm Im}\Sigma^\beta_R(p_0,p)\,,}
for real $p_0$. For fixed $m_1$ and $m_2$, these cuts exist for
\eqna\cuts
$$
\eqalignno{
u^2 -p^2&\geq (m_1+m_2)^2\,, &\cuts a\cr
u^2 -p^2&\leq (m_1-m_2)^2\,. &\cuts b\cr}
$$
The first cut is the usual zero-temperature cut corresponding to
the decay of the incoming particle, whereas the second
appears only at $T\neq 0$ and represents absorption of a particle from
the medium.
The first cut does not lend itself to non-commuting double limits, so
the only suspect is the second cut. In fact, it is this cut which is
responsible for the non-commuting double limits in the case
$m_1=m_2$\refmark{\weldonfp}. In our case however, the contribution
of this cut is perfectly well-behaved as $P^\mu\to 0$.
In fact, if we denote this contribution by $C_2(p_0,p)$, then we obtain
\eqn\secondcut{
{\rm Re}\Sigma^\beta_R(p_0,p)\ni C_2(p_0,p)={2\over\pi}
{\cal P}\int_0^{\left(p^2+(m_1-m_2)^2\right)^{1\over 2}}
 du\,u {{\rm Im}\Sigma^\beta_R(u,p)\over u^2-p_0^2}\,.}
Performing the change of variables $u\rightarrow
u/\sqrt{p^2+(m_1-m_2)^2}$, we obtain
\eqn\thirdcut{Re\Sigma^\beta_R(p_0,p)\ni C_2(p_0,p)={2\over\pi}
{\cal P}\int_0^1
 du\,u {{\rm Im}\Sigma^\beta_R(u\sqrt{p^2+(m_1-m_2)^2},p)
\over u^2-{p_0^2\over p^2+(m_1-m_2)^2}}\,.}
As long as the masses are different, the zero momentum
limit of $C_2(p_0,p)$ is well-defined and given by
\eqn\zerozero{
C_2(0,0)={2\over\pi}
\int_0^{|m_1-m_2|} du {{\rm Im} \Sigma^\beta_R(u,0)\over u}\,.}
\noindent
This limit, however, is not well-defined if the masses are equal.
Note that \zerozero\ is well-behaved, given that
${\rm Im}\Sigma^\beta(u,0)$
is odd in $u$, and goes as $u$ for small $u$.

The results of the previous section might, at first, appear to have
far-reaching consequences in the study of finite-temperature field theory,
where it has always been assumed that {\it all} one-loop self-energies
exhibit a non-analytic behavior at vanishing external four-momentum.
One may naturally wonder whether our observation has any effect on
standard computations of {\it physical} quantities, such as the
difference between Debye and plasmon masses in the standard electroweak
theory, and whether there could be any effect on studies of the
electroweak phase transition\refmark{\ewpt}. In fact it does not, as
can be argued in the following way.
Our result \pioohit\ for the $P^\mu\to 0$ limit depends on assuming
$p_0,p\ll |\Delta m^2|/T$ in \picomp, since \picomp\ is dominated
by $k\sim T$. However, the region of interest
for self-consistently finding the Debye or plasmon poles of the vector
propagator is when $p_0$ or $p$ take values of order
$m_i\gg |\Delta m^2|/T$.
In that regime, $\Delta m^2_{12}$ can be
ignored in \pik, in which case one recovers the usual non-commuting
double limits. The qualitative features of our results are shown in
\fig\qualit{For
$p_0$ and $p$ small compared to $|\Delta m^2|/T$, the functions
$\Pi^\beta_{00}(p_0,0)$ and $\Pi^\beta_{00}(0,p)$ tend to the same limit.
However, at order $m$, the functions take on different values. As the
mass difference goes to zero, it is clear that the unique limit disappears,
as well.}. For
$p_0$ and $p$ small compared to $|\Delta m^2|/T$, the functions
$\Pi^\beta_{00}(p_0,0)$ and $\Pi^\beta_{00}(0,p)$ tend to the same limit.
However, at order $m$, the functions take on different values. As the mass
difference goes to zero, it is clear that the unique limit disappears,
as well.

\appendix{A}{Polarization Tensor Calculation}

 In unitary gauge, the only one-loop momentum-dependent correction to the
photon propagator is given by fig.~2. We denote this
$\tilde\Pi_{\mu\nu}$. We will compute this contribution in the
real-time formalism, using Feynman parametrization, and
$\epsilon$-regularization of the propagators. We obtain a regulated
expression,
\eqn\pimunu{
\eqalign{\tilde\Pi_{\mu\nu}(p_0,p)= 4ie^2 M^2\int {d^4k\over (2\pi)^4}
(g_{\mu\nu}-&{k_\mu k_\nu\over M^2})\left[{\cal P}_\epsilon({1\over
D_1})
- 2\pi i\, \left(n(|k_0|)+{1\over 2}\right)\delta_\epsilon(D_1)\right]\cr
&\left[{\cal P}_\epsilon({1\over D_2}) - 2\pi i\, \left(n(|p_0+k_0|)+
{1\over2}\right)\delta_\epsilon(D_2)\right]\,,\cr }}
where $D_1=K^2-M^2$ and $D_2=(K+P)^2-m^2$, while
the regulated principal value and $\delta$-function are defined by
\eqna\deltae
$$
\eqalignno{
\delta_\epsilon(x)&={1\over \pi}{\epsilon\over x^2+\epsilon^2}\,,
&\deltae a\cr
{\cal P}_\epsilon ({1\over x}) &={x\over x^2+\epsilon^2}\,.
&\deltae b\cr}
$$

The real part of the $T\ne 0$ contribution is
\eqn\realpart{
\eqalign{Re\tilde\Pi^\beta_{\mu\nu}= 4e^2 \int {d^4k\over (2\pi)^3}
(M^2 g_{\mu\nu}-k_\mu k_\nu)& [{\cal P}_\epsilon({1\over K^2-M^2})
 n(|p_0+k_0|) \delta_\epsilon ((P+K)^2-m^2)\cr
&+{\cal P}_\epsilon ({1\over (P+K)^2-m^2})n(|k_0|)
\delta_\epsilon(K^2-M^2)] ,\cr}}
There is an implicit limit of $\epsilon\rightarrow 0$ in all the
above expressions. This limit is to be taken at the end, after all relevant
integrations.

\noindent
Let us note that
the $\epsilon$-regularization is superfluous for generic values of the
vector
and scalar masses. In the case, however, where $M=m$, the conventional
calculation becomes ambiguous at $P^\mu=0$. So, we will perform the
calculation for an arbitrary infinitesimal $\epsilon$ and we will let
$\epsilon$ go to zero at the end.

\noindent
Define
\eqn\hn{
h_n = \int dk_0 k_0^n n(|k_0|){\cal P}_\epsilon({1\over (P+K)^2-m^2})
\delta_\epsilon(K^2-M^2)}
\noindent
and
\eqn\hntilde{
\tilde h_n = \int dk_0 k_0^n n(|k_0|){\cal P}_\epsilon
({1\over (P+K)^2-M^2}) \delta_\epsilon(K^2-m^2)\,.}
In terms of these functions, the polarization tensor becomes
\eqn\pizro{
{\rm Re}\tilde\Pi^\beta_{00} = 4e^2 \int {d^3k\over (2\pi)^3}
\left [ M^2 h_0 - h_2 +M^2 \tilde h_0 - \tilde h_2 -2p_0\tilde h_1-p_0^2
\tilde h_0\right]\,.}
\noindent
The other components have similar expressions. Our task is to compute
the functions $h_n$. Let us rewrite \hn\ as
\eqn\hnmore{
\eqalign{
h_n=&{1\over \pi}\int^\infty_{-\infty}dk_0\,{k_0^n\over e^{\beta|k_0|}-1}
{\epsilon\over(k_0^2-\Omega_k^2)^2+\epsilon^2}
{(p_0+k_0)^2-\omega_{p+k}^2\over((p_0+k_0)^2-\omega_{p+k}^2)^2+
\epsilon^2}\cr
=&{1\over \pi}\int^\infty_0 {dk_0\ k_0^n\over e^{\beta k_0}-1} {\epsilon
\over
(k_0^2-\Omega_k^2)^2+\epsilon^2)} {(p_0+k_0)^2-\omega_{p+k}^2\over
((p_0+k_0)^2-\omega_{p+k}^2)^2+\epsilon^2}+(-1)^n (p_0\to -p_0)\cr
\equiv &\, H_n(p_0,p)+(-1)^n H_n(-p_0,p)\,.\cr}}
These integrals can now be calculated using the contour shown in
\fig\contour{
Contour in the complex $k^0$-plane used in the integration. The limit
$\epsilon'\to 0$ is implied.}.
Clearly, the integrals vanish along the arc. However, since
the Bose-Einstein distribution has a series of poles alond the imaginary
axis, the
integration along this axis would appear to give a non-vanishing
contribution.
It is easy to establish, however, that in the limit $\epsilon\to 0$, the term
$\delta_\epsilon$ would regulate this contribution to zero. We proceed
by expressing each fraction in the integrand as the sum or difference
of propagators with different analytic properties (i.e.\ differing in
their $i\epsilon$ prescriptions), to obtain
\eqn\exphn{
H_n(p_0,p)={i\over 4\pi}\int^\infty_0 {dk_0\ k_0^n\over e^{\beta k_0}-1}
\sum _{a,b=\pm 1}b{1\over (p_0+k_0)^2-\omega_{p+k}^2+ai\epsilon}
{1\over k_0^2-\Omega^2_k+bi\epsilon}\,.}
Next, we combine denominators using Feynman parametrization, appropriately
modified  by Weldon\refmark{\weldonfp}, and Bedaque and Das\refmark{
\bd},
\eqn\fp{
{1\over A+i\alpha\epsilon}{1\over B+i\beta\epsilon}=P\int_0^1{dx\over
\left[x(A+i\alpha\epsilon)+(1-x)(B+i\beta\epsilon)\right]^2}+2\pi i{
(\alpha-\beta)\delta(\beta A-\alpha B)\over A-B+i(\alpha-\beta)\epsilon}\,.}
We shall denote the contribution to $H_n$ from the integral over the
Feynman parameter by $H_n^x$ and the contribution from the
delta-function by $H_n^\delta$.
\eqn\hnx{
\eqalign{
H_n^x(p_0,p)={i\over 4\pi}\int^\infty_0 dk_0{k_0^n\over e^{\beta k_0}-1}
&\sum_{a,b=\pm 1}b\int_0^1 dx {1\over [(k_0+xp_0)^2-\phi_k^2
+(xa+(1-x)b)i\epsilon]^2}\cr
={i\over 4\pi}\Big\{{\partial\over\partial\epsilon}
\int^\infty_0 dk_0{k_0^n\over e^{\beta k_0}-1} &\sum _{a,b=\pm 1}b
\int^1_0 dx\, {i\over y}
\big[{1\over k_0+xp_0-\phi_k+y{i\epsilon\over 2\phi_k}}\cdot\cr
&\quad{1\over k_0+xp_0+\phi_k-y{i\epsilon\over 2\phi_k}}\big]
\Big\}_{\epsilon=0}\, ,\cr}}
where
\eqn\phik{
\phi_k=(-x(1-x)p_0^2+(1-x)\Omega^2_k+x\omega^2_{p+k})^{1\over 2}\, ,}
and
\eqn\yx{
y=x(a-b)+b\,.}
We are now able to do the $k_0$-integration by picking the poles in
the first quadrant,
\eqn\Hnx{
\eqalign{H_n^x(p_0,p)=-{i\over 2}\Big\{{\partial\over \partial\epsilon}
\big [&\int^1_0 dx
{1\over 2\phi_k+{i\epsilon\over \phi_k}}
{(\phi_k-xp_0+{i\epsilon\over 2\phi_k})^n\over e^{\beta(\phi_k-xp_0
+{i\epsilon\over 2\phi_k})}-1}\cr
+&\int^{1\over 2}_0 {dx\over 1-2x}
{1\over 2\phi_k+(1-2x){i\epsilon\over \phi_k}}
{(\phi_k-xp_0+(1-2x){i\epsilon\over 2\phi_k})^n\over e^{\beta(\phi_k-xp_0
+(1-2x){i\epsilon\over 2\phi_k})}-1}\cr
&\int^1_{1\over 2} {dx\over 1-2x}
{1\over 2\phi_k-(1-2x){i\epsilon\over \phi_k}}
{(\phi_k-xp_0-(1-2x){i\epsilon\over 2\phi_k})^n\over e^{\beta(\phi_k-xp_0
-(1-2x){i\epsilon\over 2\phi_k})}-1}\big ]\Big\}_{\epsilon=0}\,.\cr}}
Upon defining a new variable
\eqn\zk{
z_k(\alpha)=(-x(1-x)p_0^2+(1-x)\Omega_k^2+x\omega^2_{p+k}
+\alpha)^{1\over 2}\,, \qquad z_k(0)=\phi_k}
Eq.~\Hnx\ reduces to
\eqn\Hnxa{
H_n^x(p_0,p)={1\over 2}\Big\{ {\partial\over  \partial\alpha}
\int^{1\over 2}_0
{dx\over z_k}{(z_k-xp_0)^n\over e^{\beta(z_k-xp_0)}-1}\Big\}_
{\alpha=0}\,.}
The term involving the delta-function in the
Feynman parametrization formula can be easily computed to contribute
\eqn\Hndelta{
H_n^\delta(p_0,p)=-{1\over 2R}{(-p_0/2+R)^n\over e^{\beta(-p_0/2+R)}-1}
{1\over\omega^2_{p+k}-\Omega^2_k-2p_0 R}\,,}
where
\eqn\Rdef{
R=\sqrt{-{p_0^2\over 4}+{1\over 2}(\omega_{p+k}^2+\Omega^2_k)}\,.}

\noindent
The function $\tilde{h}_n$, defined by \hntilde\ can be calculated in the
same fashion.

\noindent
Recalling \pizro, and collecting only the Feynman parameter dependent terms,
we obtain
\eqn\pizrox{
\eqalign{{\rm Re}\tilde\Pi_{00}^\beta\ni
2e^2\Big\{{\partial\over \partial\alpha}
\int {d^3k\over (2\pi)^3}\int^{1\over 2}_0& dx\,
(M^2-(z_k+xp_0)^2){n(z_k+xp_0)\over z_k}\cr
+&(M^2-(\tilde z_k+(1-x)p_0)^2)
{n(\tilde z_k-xp_0)\over \tilde z_k}\Big\} _{\alpha=0}\cr
+&(p_0\to -p_0)\,,\cr}}
where $\tilde z_k\leftrightarrow z_k$ as $m\leftrightarrow M$.
The $x$-integration can be performed after a change of variables
\eqn\chv{
\eqalign{
x &\rightarrow 1-x\,\cr
\vec{k} &\rightarrow -\vec{k}-\vec{p}\,,\cr}}
in the second term, and
\eqn\w{
w=z_k+x p_0\,,}
in the resulting integrand. All the dependence on $\alpha$ resides now
in the limits of integration. Taking the derivative with respect to
$\alpha$ and the limit $\alpha\rightarrow 0$ yields,
\eqn\pioo{
\eqalign{{\rm Re}\tilde\Pi_{00}^\beta\ni 2e^2\int {d^3k\over (2\pi)^3}
\Big[
&{n(\omega_k)\over \omega_k}{M^2-(p_0-\omega_{k})^2\over
(p_0-\omega_{k})^2-\Omega_{k+p}^2}
+{n(\Omega_k)\over \Omega_k}{M^2-\Omega_k^2\over
(p_0-\Omega_k)^2-\omega_{k+p}^2}\cr
+&{1\over R}{M^2-{1\over 2}(\Omega_k^2+\omega_{p+k}^2)-p_0 R\over
\omega_{p+k}^2-\Omega_k^2 +2p_0 R}\big[{1\over e^{\beta({p_0\over 2}+R)}
-1}-
{1\over e^{\beta(-{p_0\over 2}+R)}-1}\big]\Big]\cr
+&(p_0\to -p_0)\, ,\cr}}
where $R$ is given by Eq.~\Rdef. If the delta function contribution to
the Feynman parametrization formula \fp\ were incorrectly left out, then
\pioo\ would be the complete expression for ${\rm Re}\tilde\Pi_{00}^\beta$.
It is interesting to note that in such a case \pioo\ would agree with the
expression derived within
the imaginary-time formalism for $p_0=2\pi i\ell T$, since then the third
term would vanish identically. However, \pioo\ as it stands is the wrong
analytic continuation of the euclidean expression. Indeed, the
contribution of the delta function \Hndelta\ {\it exactly} cancels the
third term in \pioo\ and, therefore, the complete expression is
\eqn\picomp{
\eqalign{{\rm Re}\tilde\Pi_{00}^\beta=4e^2\int {d^3k\over (2\pi)^3}
\Big[
&{n(\omega_k)\over 2\omega_{k}}{M^2-(p_0-\omega_{k})^2\over
(p_0-\omega_{k})^2-\Omega_{k+p}^2}
+{n(\Omega_k)\over 2\Omega_k}{M^2-\Omega_k^2\over (p_0-\Omega_k)^2
-\omega_{k+p}^2}\Big]\cr
+&(p_0\to -p_0)\, ,\cr}}
which agrees with the imaginary-time expression for real $p_0$.

P.A.\ was supported by DOE grant DE-FG06-91ER40614. P.B.\ was
supported in part by DOE grant DE-FG02-91ER40685 and in part by CAPES.
A.D.\ was supported in part by DOE grant DE-FG02-91ER40685. A.D.\ also
thanks the Division of Educational Programs of Argonne National
Laboratory. S.V.\ was supported by DOE grants
DE-FG06-91ER40614 and W-31-109-ENG-38.
S.V.\ acknowledges useful discussions with G.~Bodwin, L.~Brown,
I.~Knowles, L.~Yaffe, and C.~Zachos and thanks U.~Sarid for his
Feynman diagram drawing code.

\listrefs
\listfigs

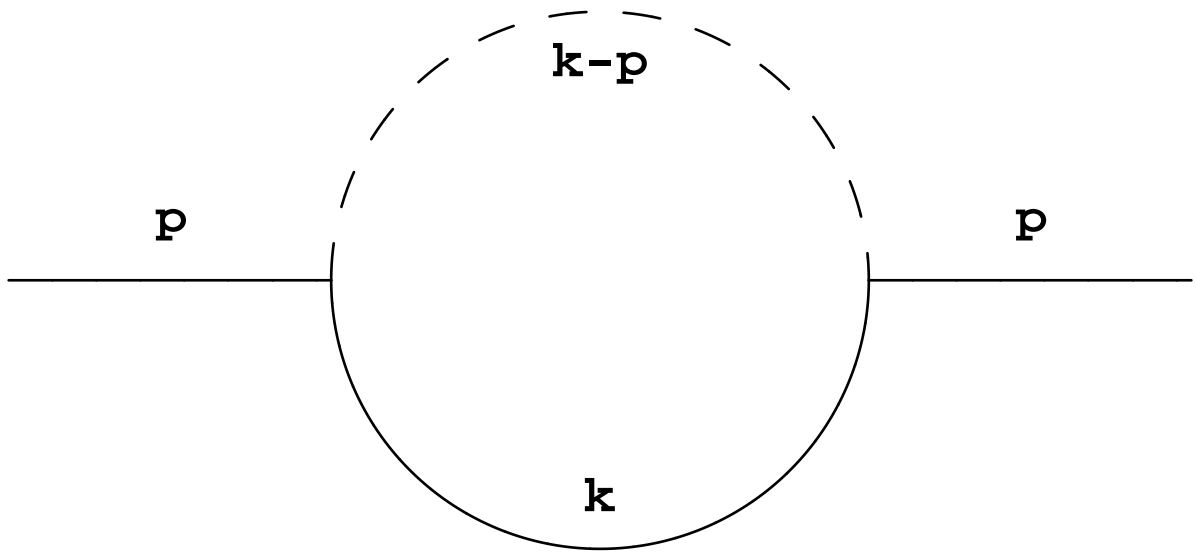
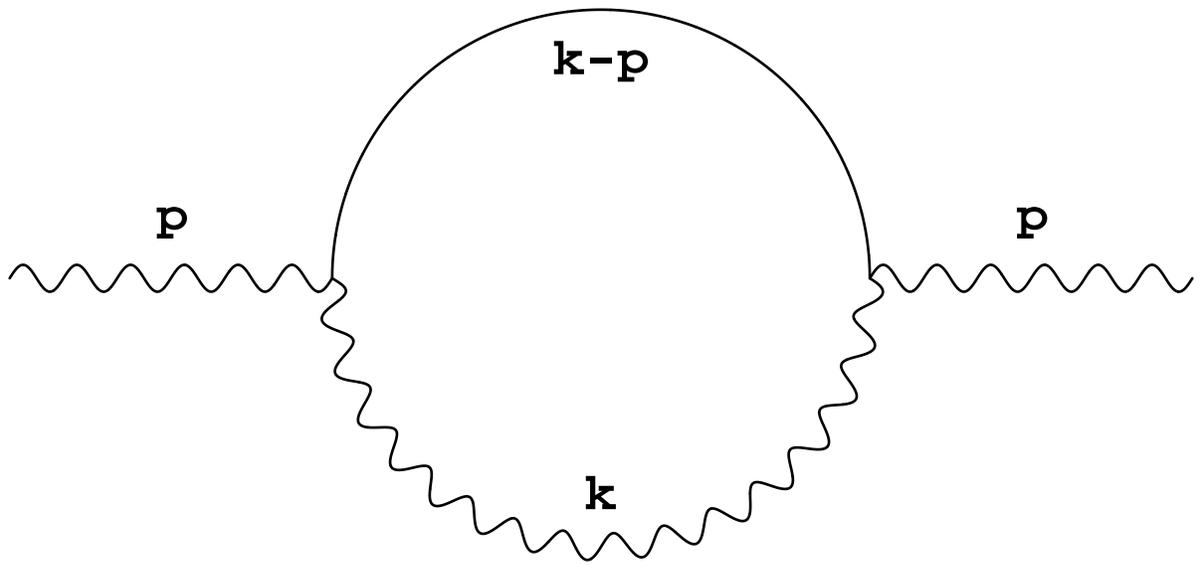
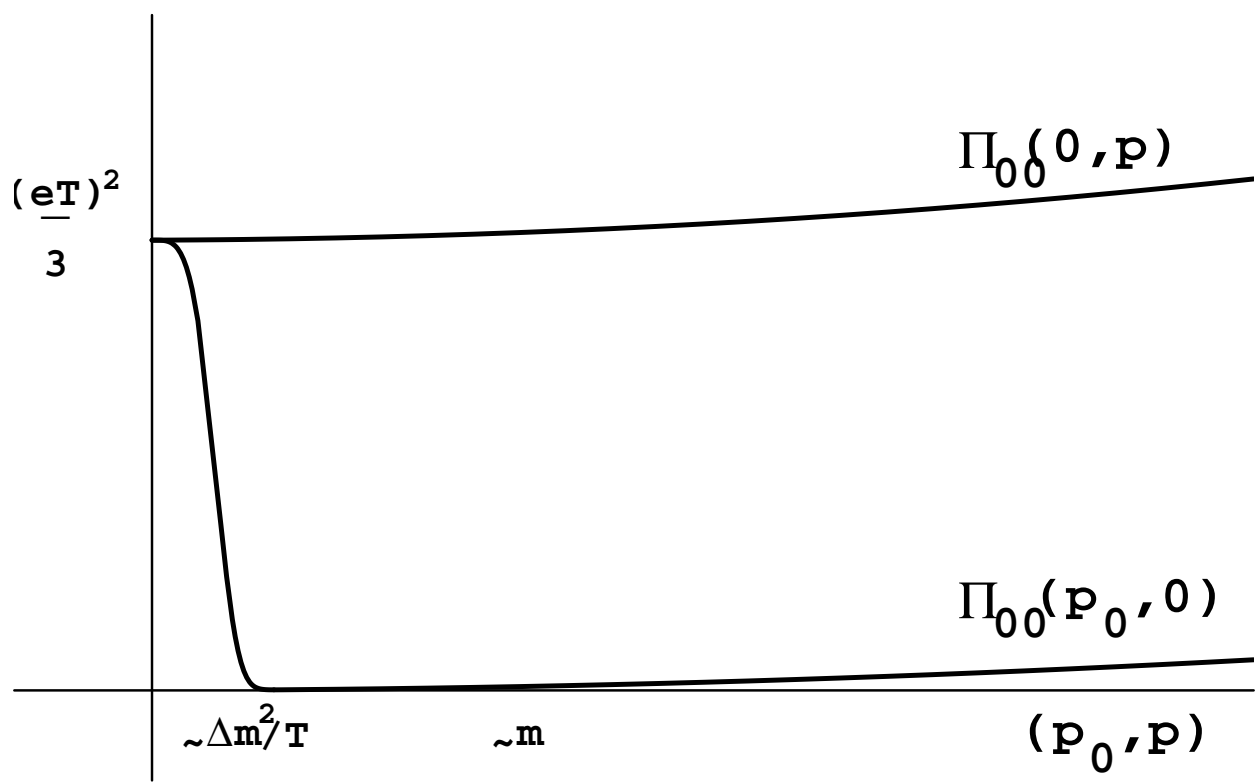
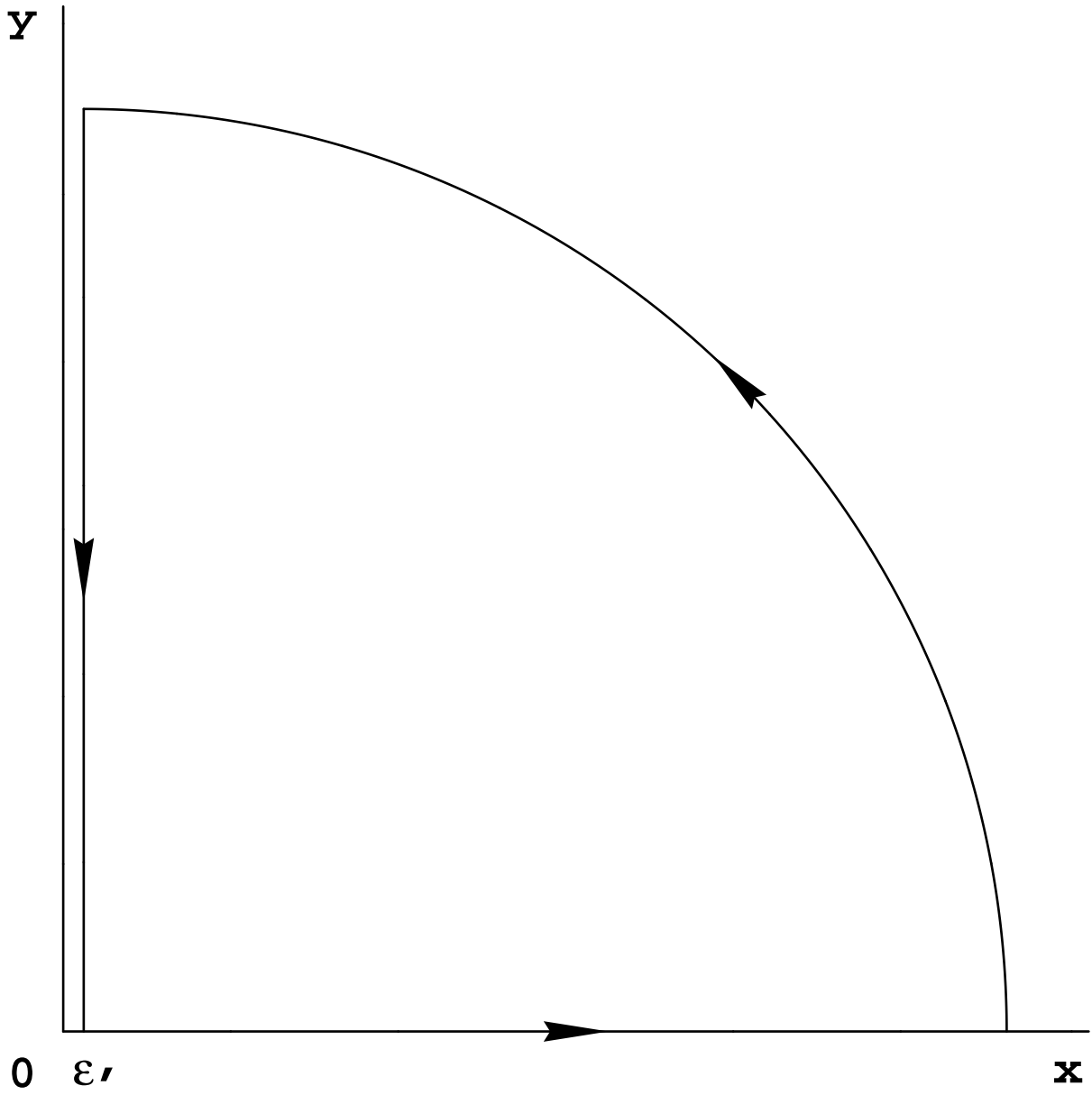
\end